\documentstyle[12pt]{article}

\begin{document}

\begin{center}
\baselineskip 40pt

\vskip 2cm

{\Large {\bf Quantum Zeno and Anti-Zeno Effect without Rotating
Wave Approximation }}

\vskip 1cm

{\large H. Zheng$^1$ and S. Y. Zhu$^2$}

{$^1$Department of Physics, Shanghai Jiao Tong University,
Shanghai 200030, China}

{$^2$Department of Physics, Hong Kong Baptist University, Hong
Kong, China}

{\bf Abstract}
\end{center}

\baselineskip 20pt

The effect of the anti-rotating terms on the short-time evolution
and the quantum Zeno (QZE) and anti-Zeno (AQZE) effects is studied
for a two-level system coupled to a bosonic environment. A unitary
transformation and perturbation theory are used to obtain the
electron self-energy, energy shift and the enhanced QZE or the
AQZE, simultaneously. The calculated Zeno time depends on the
atomic transition frequency sensitively. When the atomic
transition frequency is smaller than the central frequency of the
spectrum of boson environment, the Zeno time is prolonged and the
anti-rotating terms enhance the QZE; when it is larger than that
the Zeno time is reduced and the anti-rotating terms enhance the
AQZE.

\vskip 1cm


\pagebreak

\baselineskip 20pt

The quantum Zeno effect (QZE) and anti-Zeno effects (AQZE) have
been widely discussed for decades theoretically\cite{zeno} and
recently experimentally\cite{exp}. In an unstable quantum
two-level (multi-level) system, frequently measurement can reduce
or accelerate the decay processes
\cite{zeno,kk,fac1,fac2,fac3,yz}. The survival probability at an
excited state of a quantum system interacting with an environment
is a decaying function of time. Frequent measurements at extremely
short time interval may slow down the decay process, because the
decay of the excited state is almost zero at the beginning of the
decay process\cite{zeno,kk}, which is known as the QZE. It was
also found that if the measurement time interval is short, but not
extremely short, the decay of the excited state could be
accelerated\cite{kk}, which is known as the AQZE. Let $P(\tau)$
denote the survival probability (after a short time interval
$\tau$) at the initial state, which can be written as
$P(\tau)=\exp(-\gamma(\tau)\tau)$. After $N$ time measurements at
equal $\tau$, the survival probability reads
$P^N(\tau)=\exp[-\gamma(\tau)N\tau]=\exp[-\gamma(\tau)t]$ with
$\gamma(\tau)$ the effective decay rate. If $N=1$,
$P(t)=\exp(-\gamma(t)t)$, which goes to $P(t)\to\exp(-\gamma_0t)$
for large enough $t$, where $\gamma_0$ is the decay rate under the
Weisskopf-Wigner approximation. We will have the QZE if
$\gamma(\tau)<\gamma_0$, and the AQZE if $\gamma(\tau)>\gamma_0$.

It is well known that the whole spectrum of the environment at
off-resonance (with the transition frequency) is important for the
QZE and AQZE. If the atomic transition frequency is located not at
the maximum of the spectrum, we can have the QZE and AQZE
depending on the measurement time interval. In the previous
studies on QZE and AQZE \cite{zeno,kk,fac1,fac2,fac3,yz}, the
rotating wave approximation (RWA) is used. However, this
approximation raises the question what the influence of the
anti-rotating terms on the QZE and AQZE is, as they might have the
same order contribution as the spectrum components off-resonant
with the atomic transition, especially for the QZE where
measurement time interval is extremely short\cite{zeno,kk,fac1}.
Therefore we ask ourselves, "what is the role of anti-rotating
terms on the QZE and AQZE?"

The model to describe an unstable quantum system is the following
spin-boson model with the Hamiltonian\cite{rmp,scu},
\begin{eqnarray}
&&H={\frac{1}{2}}\omega_0 \sigma _{z} +\sum_{k}\omega _{k}b_{k}^{\dag }b_{k} +%
{\frac{1}{2}}\sum_{k}g_{k}(b_{k}^{\dag }+b_{k})\sigma _{x},
\end{eqnarray}
$b_{k}^{\dag }$ ($b_{k}$) is the creation (annihilation) operator
of boson mode with frequency $\omega _{k}$, $\sigma _{x}$ and
$\sigma _{z}$ are Pauli matrices describing the two-level system.
$\omega_0 $ is the transition frequency between the up $|+\rangle$
and down state $|-\rangle$: $\sigma_z|\pm\rangle=\pm|\pm\rangle$.
$g_{k}$ is the coupling between the two-level system and the
environment, which can be characterized by the interacting
spectrum\cite{kk,fac1,rmp}: $G(\omega)={1\over
4}\sum_{k}g^2_{k}\delta(\omega-\omega_k)$. This model Hamiltonian
is used for a large number of different physical and chemical
processes, such as the atom-field interaction and the QZE in
quantum optics\cite{kk,fac1,scu}, coupled quantum dots on a solid
state substrate\cite{rmp,sbm}, and the macroscopic quantum
coherence experiment in SQUID's\cite{dos1,dos2}.

The Hamiltonian (1) cannot be solved exactly and usually the RWA
is used\cite{kk,fac1,scu} for which the Hamiltonian is taken to be
\begin{eqnarray}
&&H_{RWA}={\frac{1}{2}}\omega_0 \sigma _{z} +\sum_{k}\omega _{k}b_{k}^{\dag }b_{k} +%
{\frac{1}{2}}\sum_{k}g_{k}(b_{k}^{\dag }\sigma_{-}+b_{k}\sigma
_{+}),
\end{eqnarray}
where $\sigma_{\pm}={1\over 2}(\sigma_x\pm i\sigma_y)$. $H_{RWA}$
can be solved in the so-called one-boson sector with the initial
state $|\psi(0)\rangle=|+\rangle |\{0_k\}\rangle$, where
$|\{0_{k}\}\rangle$ is the vacuum state for every $k$. The
survival amplitude of finding the system still in
$|\psi(0)\rangle$ at $\tau>0$ is\cite{kk,fac1}
\begin{eqnarray}
&&x_{RWA}(\tau)=\frac{1}{2\pi
i}\int_B\frac{e^{p\tau}dp}{p+i\omega_0+\frac{1}{4}\sum_k
\frac{g^2_k}{p+i\omega_k}},
\end{eqnarray}
where $B$ is the so-called Bromwich path. The survival probability
in the initial state is $P_{RWA}(\tau)=|x_{RWA}(\tau)|^2$ and the
effective decay rate $\gamma_{RWA}(\tau)$ for a short interval
$\tau$ can be calculated as follows\cite{kk},
\begin{eqnarray}
&&\gamma_{RWA}(\tau)=2\pi\int^{\infty}_0d\omega G(\omega)F(\omega-\omega_0),\\
&&F(\omega-\omega_0)=2\sin^2\left[\frac{\omega-\omega_0}{2}\tau\right]/
\pi\tau(\omega-\omega_0)^2.
\end{eqnarray}
Since $F(\omega-\omega_0)\to \delta(\omega-\omega_0)$ (the Dirac
$\delta-$function) for large enough $t$, we have the decay rate
$\gamma_0=2\pi G(\omega_0)$ in the Weisskopf-Wigner approximation.
Eqs.(3) and (4) are main results of the RWA.

When the anti-rotating terms are included, the above method is no
longer valid. Here we present an analytical approach, based on
unitary transformation and perturbation theory to calculate the
survival amplitude and the effective decay rate for Hamiltonian
(1) in order to clarify the impact of the anti-rotating terms on
the short time evolution and on the QZE and AQZE. In the following
the interacting spectrum of the environment is assumed as,
\begin{eqnarray}
&&G(\omega)=\frac{{1\over
2}\alpha\omega\Omega^4}{(\omega^2-\Omega^2)^2+\Gamma^2\omega^2},
\end{eqnarray}
where coupling strength, $\alpha$, is a dimensionless constant and
$\Omega$ is the center of gravity of the spectrum\cite{dos1,dos2},
and $\Gamma$ is the width of the spectrum. We will show that the
off-resonance ratio $\omega_0/\Omega$ plays an important role and
the effect of the anti-rotating terms must be taken into account
especially for the off-resonant case $\omega_0/\Omega\ll
(1-\Gamma/\Omega)$. We note that when $\Gamma^2\ll\Omega^2$,
$G(\omega)$ is mainly a sharp Lorentzian-type peak similar to the
case of resonant Rabi oscillation\cite{kk,dos1,dos2}. Throughout
this paper we set $\hbar =1$.

We treat the anti-rotating terms by a unitary
transformation\cite{zhe}: $H^{\prime }=\exp (S)H\exp (-S)$ with
\begin{equation}
S=\sum_{k}\frac{g_{k}}{2\omega _{k}}\xi _{k}(b_{k}^{\dag
}-b_{k})\sigma _{x} .
\end{equation}
Here we introduce in $S$ a $k$-dependent function $\xi _{k}$ and
its form will be determined later. The transformation can be
carried out, and the result is $H^{\prime }=H_{0}^{\prime
}+H_{1}^{\prime }+H_{2}^{\prime }$,
\begin{eqnarray}
&&H_{0}^{\prime }={\frac{1}{2}}\eta \omega_0 \sigma _{z}
+\sum_{k}\omega _{k}b_{k}^{\dag }b_{k}
-\sum_{k}\frac{g_{k}^{2}}{4\omega _{k}}\xi _{k}(2-\xi
_{k}) , \\
&&H_{1}^{\prime }={\frac{1}{2}}\sum_{k}g_{k}(1-\xi _{k})(b_{k}^{\dag
}+b_{k})\sigma _{x} -{\frac{1}{2}}\eta \omega_0 i\sigma_y\sum_{k}\frac{g_{k}}{%
\omega _{k}} \xi _{k}(b_{k}^{\dag }-b_{k}), \\
&&H_{2}^{\prime }={\frac{1}{2}}\omega_0\sigma_z\left( \cosh \{\sum_{k}\frac{%
g_{k}}{\omega _{k}}\xi _{k}(b_{k}^{\dag }-b_{k})\}-\eta \right)  \nonumber \\
&&-{\frac{1}{2}}\omega_0 i\sigma_y\left( \sinh
\{\sum_{k}\frac{g_{k}}{ \omega
_{k}}\xi _{k}(b_{k}^{\dag }-b_{k})\}-\eta \sum_{k}\frac{g_{k}}{ \omega _{k}}%
\xi _{k}(b_{k}^{\dag }-b_{k})\right)
\end{eqnarray}
where
\begin{eqnarray}
&&\eta =\exp [-\sum_{k}\frac{g_{k}^{2}}{2\omega _{k}^{2}}\xi
_{k}^{2}].
\end{eqnarray}
$H^{\prime}_0$ is the unperturbed part of $H^{\prime}$ and,
obviously, it can be solved exactly because in which the two-level
system and the bosons are decoupled. The eigenstate of
$H_{0}^{\prime}$ is a direct product: $|\pm\rangle
|\{n_{k}\}\rangle$, where $|\{n_{k}\}\rangle$ is the eigenstate of
bosons with $n_{k}$ bosons for mode $k$. In particular, the ground
state of $H_{0}^{\prime}$ is $|g_{0}\rangle =|-\rangle
|\{0_{k}\}\rangle$.

$H_{1}^{\prime }$ and $H_{2}^{\prime }$ depend on $g_k$ and are
small, which are treated as perturbation. Because of the
definition of $\eta$ in Eq.(11), $H'_2$ contains the terms of
two-boson and multi-boson non-diagonal transitions and its
contribution to physical results is $(g^2_k)^2$ and higher. So,
$H'_2$ can be omitted and we approximate $H'\approx H'_0+H'_1$.
$H'_1$ contains the terms of single-boson transition and we chose
$\xi_k$ as
\begin{eqnarray}
&&\xi _{k}=\frac{\omega _{k}}{\omega _{k}+\eta\omega_0 },
\end{eqnarray}
so that $H'_1$ is of the form
\begin{eqnarray}
&&H_{1}^{\prime }=\eta \omega_0 \sum_{k}\frac{g_{k}}{\omega _{k}}%
\xi _{k}\left[ b_{k}^{\dag }\sigma _{-}+b_{k}\sigma _{+}\right],
\end{eqnarray}
It is easily to check that $H_{1}^{\prime }|g_{0}\rangle =0$. We
note that the transformed Hamiltonian $H'$ is of a form similar to
$H_{RWA}$ but $\omega_0$ and $g_k/2$ in (2) are replaced by
$\eta\omega_0$ and $\eta\omega_0g_k\xi_k/\omega_k$, respectively.

As $H'_1|g_0\rangle=0$, we have $H'|g_0\rangle\approx
(H'_0+H'_1)|g_0\rangle=E_g|g_0\rangle$ with the ground state
energy $E_g=-{1\over 2}\eta\omega_0
-\sum_{k}\frac{g^2_k}{4\omega_k}\xi_k(2-\xi_k)$, which is lower
than the ground state energy $E^{RWA}_g=-{1\over 2}\omega_0$ in
the RWA . It can be seen that the third term in $H'_0$ (or the
second term in $E_g$) does not depend on the atomic transition
frequency; it is the self-energy of the free electron due to the
vacuum fluctuations. In previous treatment\cite{scu} the
self-energy usually needs to be calculated separately.

The transition frequency (originally $\omega_0$) is modified to
$\omega_a=\eta\omega_0$ with the modification factor $\eta$ due to
the anti-rotating terms. Figure 1 shows $\eta$ as a function of
$\alpha$ for various values of $\omega_0$. The horizontal dotted
line is $\eta=1$ in the RWA. We note that $(\eta-1)\omega_0$ is an
energy shift of the two-level system resulted from the
anti-rotating terms. By using this unitary transformation Eq.(7),
we can obtain the self-energy and the energy shift simultaneously.

Since $|g_0\rangle$ is the ground state of $H'$, the ground state
of the original $H$ is $\exp(-S)|g_0\rangle$. Then, the survival
amplitude of finding the system in the initial state is
$x(\tau)=\langle +|\langle \{0_k\}|\exp(-iH'\tau)|+\rangle
|\{0_k\}\rangle$. Since $H'$ is of a form similar to $H_{RWA}$,
$x(\tau)$ can be calculated in the so-called one-boson sector in
the same way as $x_{RWA}(\tau)$ was,
\begin{eqnarray}
&&x(\tau)=\frac{1}{2\pi
i}\int_B\frac{e^{p\tau}dp}{p+i\eta\omega_0+\sum_k
\frac{V^2_k}{p+i\omega_k}},
\end{eqnarray}
where $V_k=\eta\omega_0g_k\xi_k/\omega_k$. The survival
probability in the initial state is $P(\tau)=|x(\tau)|^2$ and the
effective decay rate $\gamma(\tau)$ for a short interval $\tau$ is
\begin{eqnarray}
&&\gamma(\tau)=2\pi\int^{\infty}_0d\omega
G'(\omega)F(\omega-\eta\omega_0),
\end{eqnarray}
where
$G'(\omega)=4G(\omega)(\eta\omega_0)^2/(\omega+\eta\omega_0)^2=G(\omega)f(\omega)$.
The spectrum is modulated by the factor
$f(\omega)=(2\omega_a)^2/(\omega+\omega_a)^2$. The physics of this
factor is clear. It is proportional to $1/(\omega+\omega_a)^2$
because it comes from the anti-rotating terms. It is equal to 1
for $\omega=\omega_a$, because the decay rate at large enough time
is proportional to $G(\omega_a)$. Please note that $f(\omega)>1$
(or $<1$) and $G'(\omega)>$(or $<$)$G(\omega)$ for $\omega<$(or
$>$)$\omega_a$.

For the short-time limit the survival probability is quadratic in
$\tau$: $P(\tau)=1-\tau^2/\tau^2_Z$ for $\tau\to 0$, which is
explicitly different from the exponential decay
law\cite{kk,fac1,fac3}. The quantity $\tau_Z$ is referred to as
"Zeno time"\cite{fac1,fac3} and can be calculated by using
Eq.(15),
\begin{eqnarray}
&&\tau_Z=\left.\left(\frac{d}{d\tau}\gamma(\tau)\right)^{-1/2}\right|_{\tau\to
0}=\left(\int^{\infty}_0 d\omega G'(\omega)\right)^{-1/2},
\end{eqnarray}
while the Zeno time in the RWA is approximately
$\tau^{RWA}_Z=(\int^{\infty}_0 d\omega G(\omega))^{-1/2}$. One can
check that $\tau^{RWA}_Z$ appears to be independent of $\omega_0$
because the integrand $G(\omega)$ does not depend on $\omega_0$.
This should not be physically correct since the short-time
evolution $P(\tau)=1-\tau^2/\tau^2_Z$ should depend on where the
transition frequency $\omega_0$ is located in the interacting
spectrum. Our $\tau_Z$ without RWA does depend on $\omega_0$,
since $G'(\omega)$ is a function of $\omega_0$. When the atomic
transition frequency is smaller than the central frequency of the
spectrum the Zeno time is prolonged; when it is larger the Zeno
time is reduced, as shown in Fig.2. Also the energy shift is
dependent on the location of $\omega_0$ in the spectrum. The
smaller the ratio $\omega_0/\Omega$ or the larger the $\alpha$
(stronger interaction), the larger the energy shift will be (see
Fig. 1).

In Fig.3, $\gamma(\tau)$ is plotted for $\omega_0$ located in the
low frequency part of the spectrum, $\omega=0.2\Omega$. The dashed
line is the result of RWA, and one can see that for extremely
short time ($\gamma_0\tau<0.01$) RWA predicts the QZE but for a
short time ($\gamma_0\tau>0.01$) there is a possibility of AQZE.
However, by taking into account the anti-rotating terms, we only
have the QZE and no AQZE. This is a general conclusion of our
calculation for the off-resonant spectrum with $\omega_0/\Omega\ll
(1-\Gamma/\Omega)$, which is different from that of Ref.\cite{kk}.

Kofman and Kurizki\cite{kk} concluded that when $\omega_0$ is
significantly detuned from the maximum of $G(\omega)$ at $\Omega$
(so that $G(\omega_0)\ll G(\Omega)$), the effective decay rate
$\gamma(\tau)$ grows with decreasing $\tau$ and leads to the AQZE
of decay acceleration by frequent measurements. The reason can be
understood by checking $G(\omega)$ and $G'(\omega)$ in Eqs.(4) and
(15). The modification factor $f(\omega)$ due to the anti-rotating
terms on the spectrum $G(\omega)$ is $f(\omega)<1$ for
$\omega>\omega_a$. When $\omega_a$ is much smaller (or smaller)
than the spectrum center frequency $\Omega$, the spectrum
$G'(\omega)$ is greatly flattened (or flattened) compared with
$G(\omega)$, see the inset of Fig.3 (or Fig. 4). The dephasing
function\cite{kk} $F(\omega-\omega_0)$ is mainly a single-peak
function with peak at $\omega_0$ and width $\sim 1/\tau$. Since
the integrand in Eq.(4) is $G(\omega)F(\omega-\omega_0)$, when
$\omega_0$ is far below the maximum of $G(\omega)$, one can check
that $\gamma_{RWA}(\tau)$ grows with decreasing $\tau$ (AQZE)
because $F(\omega-\omega_0)$ is then probing more of the rising
part of $G(\omega)$. Our result is different from Ref.\cite{kk}
because the integrand in Eq.(15) is $G'(\omega)F(\omega-\omega_a)$
and $\gamma(\tau)$ decreases with decreasing $\tau$ (QZE) since
$F(\omega-\omega_a)$ already covers the main part of $G'(\omega)$.

In Fig.4 we plot $\gamma(\tau)$ for $\omega_0=0.5\Omega$. Here RWA
predicts anti-QZE for $\gamma_0\tau>0.09$, but our result predicts
very weak anti-QZE for the region $0.24<\gamma_0\tau<0.58$,
because the factor $f(\omega)$ coming from the anti-rotating terms
slightly flattens the spectrum (see the inset of Fig.4).

For the resonant case, $\omega_a\sim\Omega$, our calculations with
anti-rotating terms are nearly the same as those of RWA, because
the factor $f(\omega)$ come from the anti-rotating terms is almost
equal to 1 around the frequency $\omega=\Omega\sim\omega_a$, and
changes the spectrum very little. This can be understood easily
since the RWA is a good approximation for the resonant case.

Furthermore, for $\omega_0$ larger than $\Omega$, we have
$f(\omega)>1$ for $\omega < \omega_0$, that is to say, the main
part of the spectrum enhanced by the anti-rotating terms. In Fig.5
we plot $\gamma(\tau)$ for $\omega_a=1.5\Omega$, the region for
AQZE is wider than predicted by RWA, because the anti-rotating
terms raise the peak of $G'(\omega)$; see the inset of Fig.5.

In summary: The impact of the anti-rotating terms on the
short-time evolution and the quantum Zeno and anti-Zeno effects is
studied for a two-level system coupled to a bosonic environment.
We present an analytical approach, based on a unitary
transformation and a perturbation method. With this method, we can
simultaneously obtain the electron self-energy, energy shift and
the enhancement of the quantum Zeno or the anti-Zeno effects. The
effective decay rate is calculated. The Zeno time depends on the
atomic transition frequency sensitively. When the atomic
transition frequency is smaller than the central frequency of the
spectrum the Zeno time is prolonged and the anti-rotating terms
enhance the QZE; when it is larger than that the Zeno time is
reduced and the anti-rotating terms enhance the AQZE.

\vskip 0.5cm

{\noindent {\large {\bf Acknowledgement}}}

\bigskip

Part of the work of H.Z. was done when he visited Department of
Physics, Baptist University of Hong Kong. H.Z. was supported
partly by the National Natural Science Foundation of China (Grants
No.90503007 and No.10734020). S.Y.Z. was supported by
HKUST/CA/06-07/02 and FRG of HKBU.

{\rm \baselineskip 20pt }

{\rm \newpage }

\begin{center}
{\rm {\Large {\bf Figure Captions }} }
\end{center}

{\rm \vskip 0.5cm }

{\rm \baselineskip 20pt }

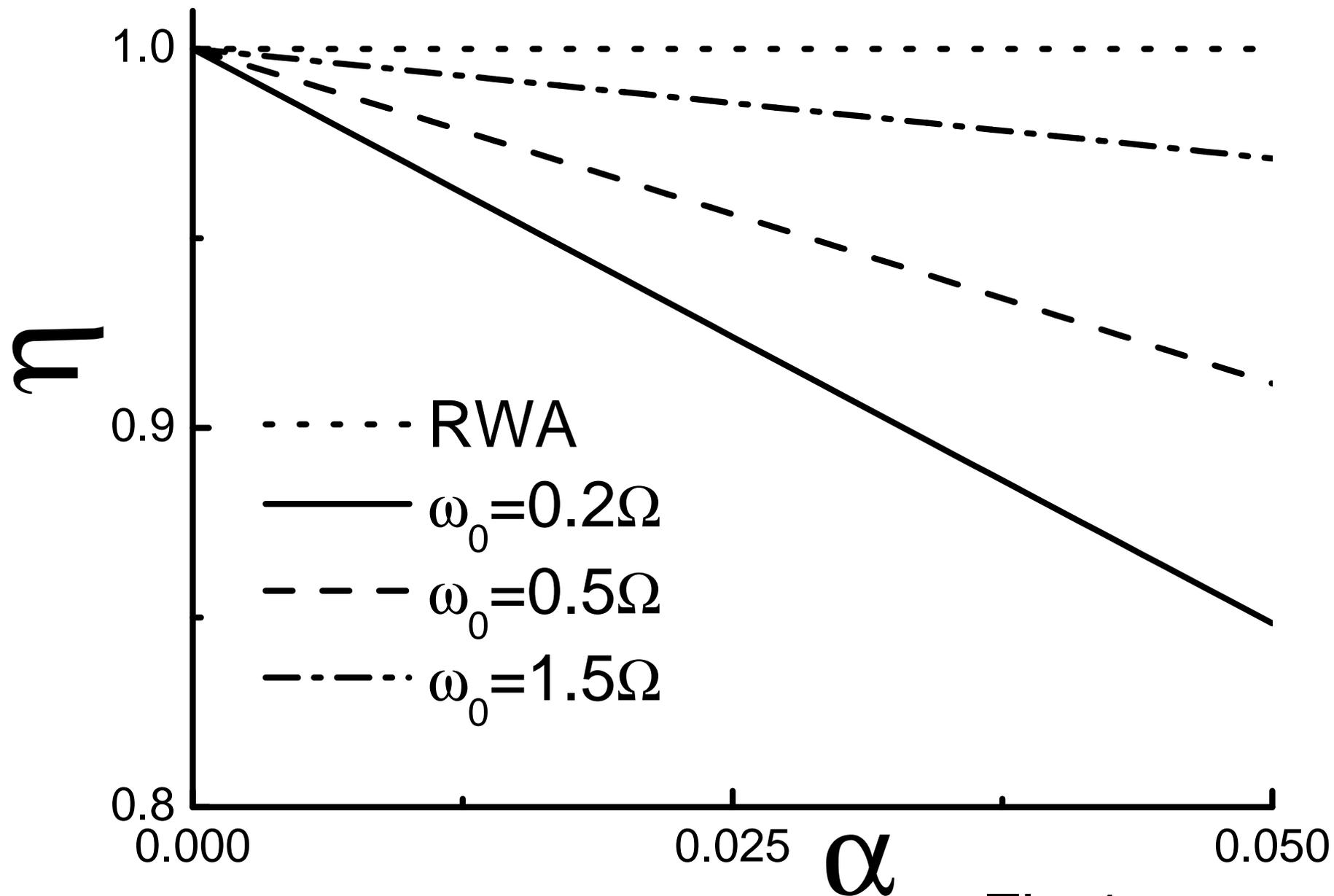
{\rm {\bf Fig.1}~~~The renormalization factor $\eta$ as a function
of the coupling $\alpha$.

{\rm \vskip 0.5cm }

{\rm {\bf Fig.2}~~~The Zeno time as a function of the transition
frequency $\omega_0$.

{\rm \vskip 0.5cm }

{\rm {\bf Fig.3}~~~Decay rate $\gamma(\tau)$ for $\alpha=0.02$,
$\omega_0=0.2\Omega$ and $\Gamma=0.4\Omega$. The dashed line is
the result of RWA. Inset: $G(\omega)$ (dotted line) and
$G'(\omega)$ (solid line).

{\rm \vskip 0.5cm }

{\rm {\bf Fig.4}~~~ Decay rate $\gamma(\tau)$ for the below
resonant spectrum with $\alpha=0.02$, $\omega_0=0.5\Omega$ and
$\Gamma=0.4\Omega$. The dashed lines is the result of RWA. Inset:
$G(\omega)$ (dotted line) and $G'(\omega)$ (solid line)

{\rm \vskip 0.5cm }

{\rm {\bf Fig.5}~~~Decay rate $\gamma(\tau)$ for the above
resonant spectrum with $\alpha=0.02$, $\omega_0=1.5\Omega$ and
$\Gamma=0.4\Omega$. The dashed line is the result of RWA. Inset:
$G(\omega)$ (dotted line) and $G'(\omega)$ (solid line)

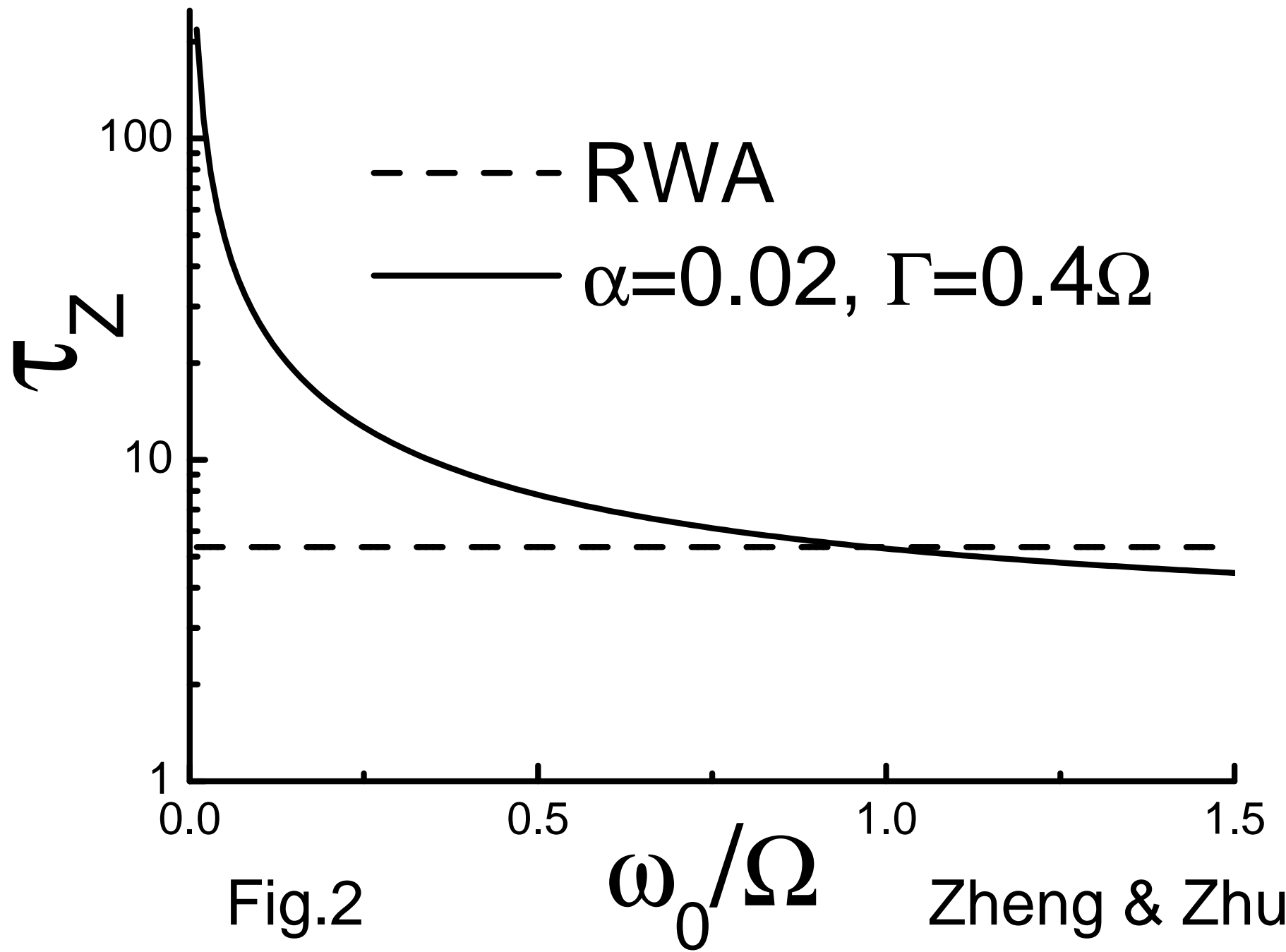
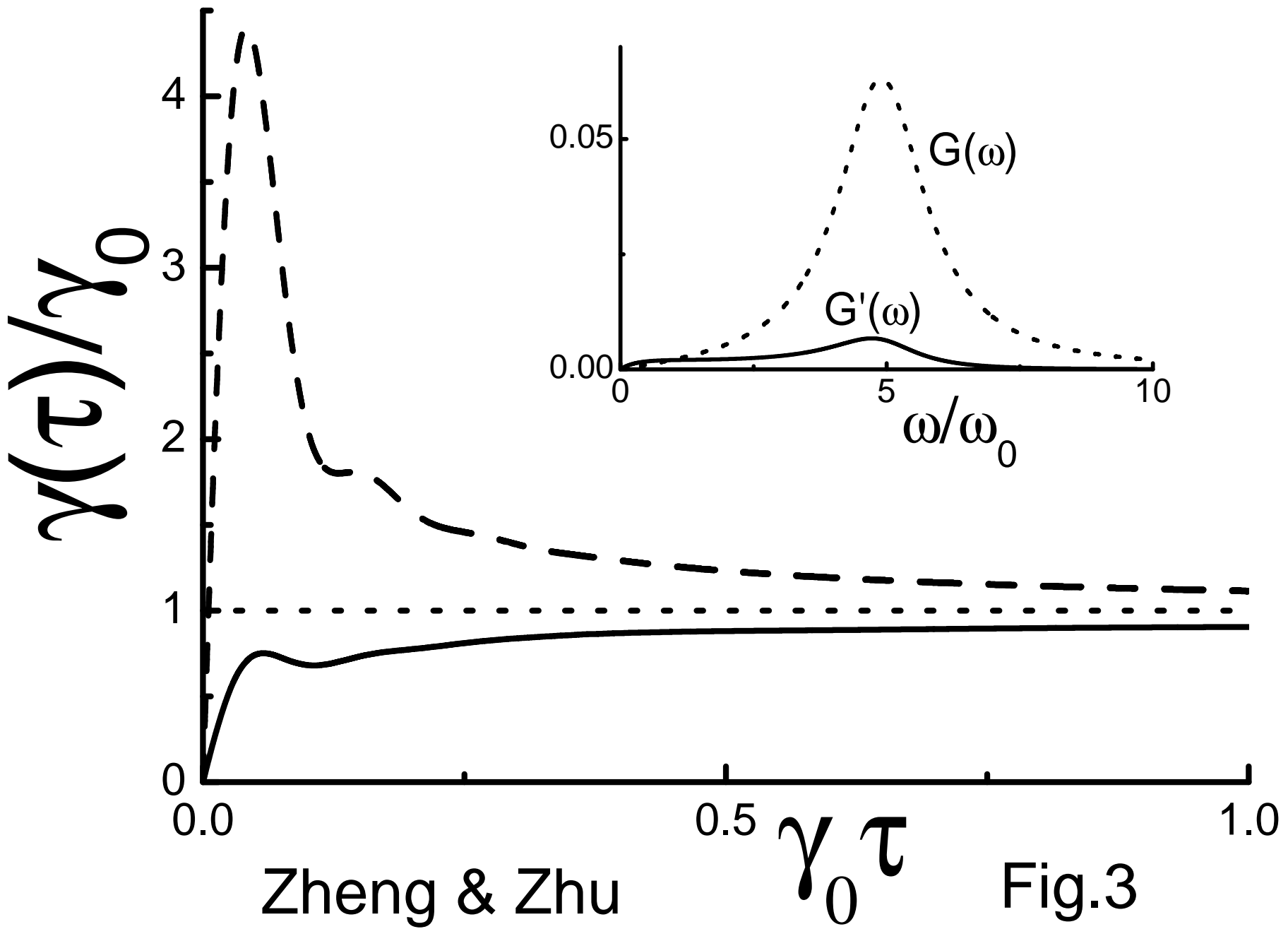
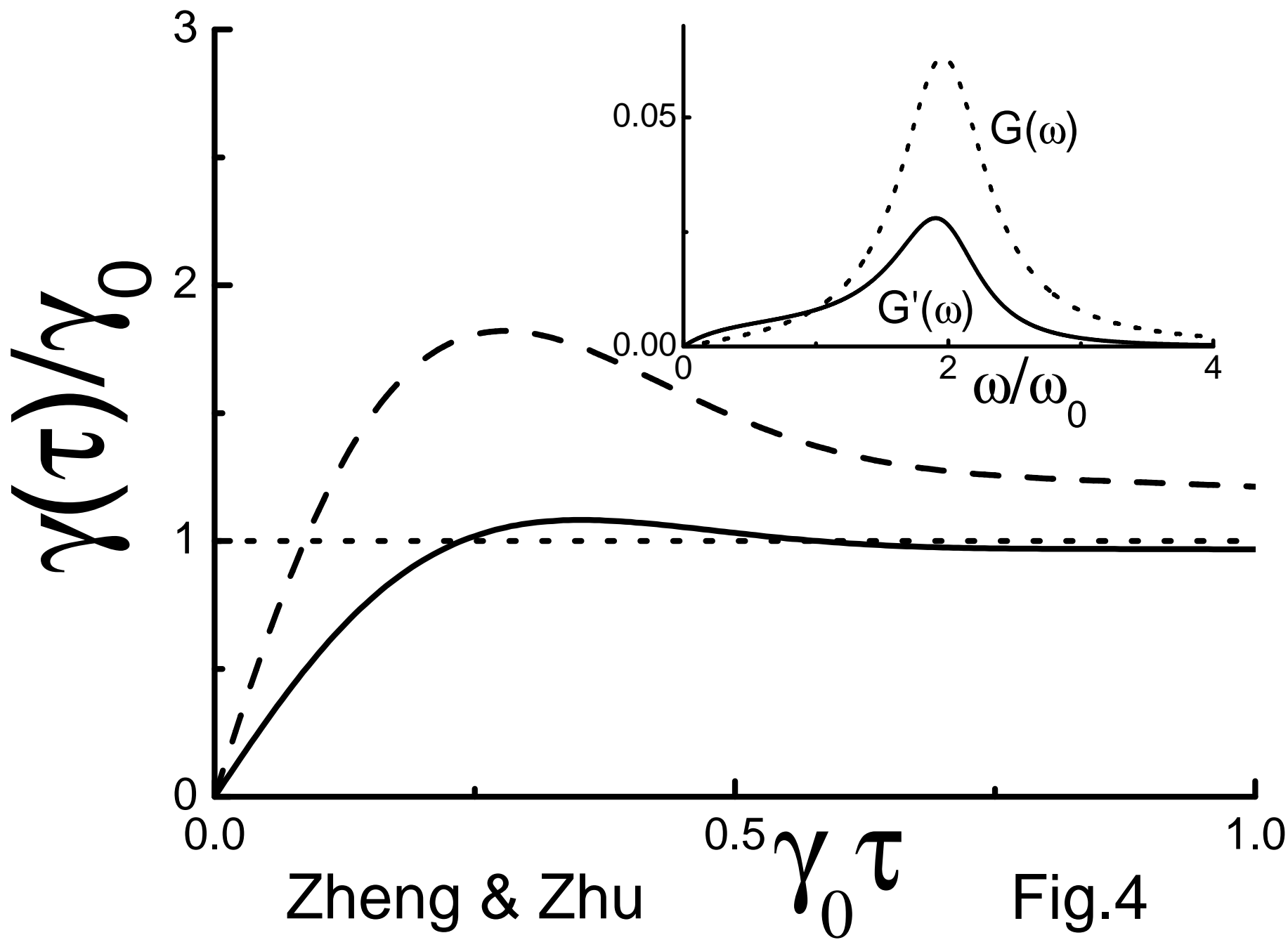
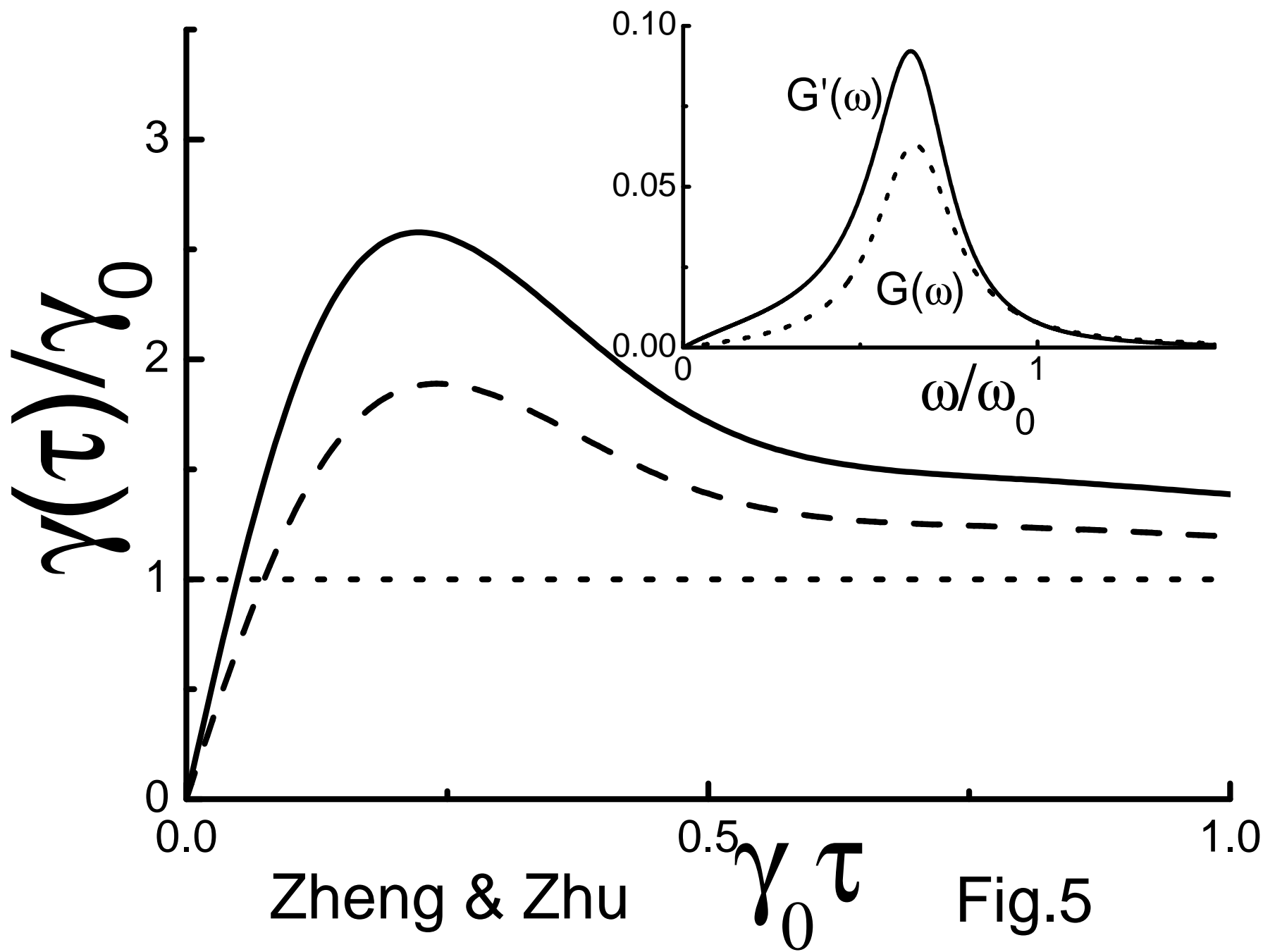
\end{document}